\documentclass[aps,12pt,superscriptaddress,preprintnumbers,
                secnumarabic,nofootinbib]{revtex4}

\usepackage{epsfig}
\usepackage{bbm}

\newcommand{\lbfig}[1]{\refstepcounter{fig} \label{#1} }
\newcounter{fig}

\begin{document} 

\preprint{SPIN-06-11, ITP-UU-06-13}

\title{\Large A solution to the cosmological constant problem}

\author{Tomislav Prokopec
       }


\email[]{T.Prokopec@phys.uu.nl}
\affiliation{Institute for Theoretical Physics (ITP) \& Spinoza Institute,
             Utrecht University, Leuvenlaan 4, Postbus 80.195, 
              3508 TD Utrecht, The Netherlands}

\begin{abstract}
 We argue that, when coupled to Einstein's theory of gravity, the Yukawa
theory may solve the cosmological constant problem in the following sense:
The radiative corrections of fermions generate an effective potential
for the scalar field, such that the effective cosmological
term $\Lambda_{\rm eff}$ is dynamically driven to zero. Thence,
for any initial positive cosmological constant
$\Lambda_0$, $\Lambda_{\rm eff} = 0$ is an attractor 
of the semiclassical Einstein theory coupled to fermionic and scalar matter
fields. When the initial cosmological term is negative, 
$\Lambda_{\rm eff}=\Lambda_0$ does not change. 
Next we argue that the dark energy of the Universe
may be explained by a GUT scale fermion with a mass, 
$m\simeq
 4.3\times 10^{15}~{\rm GeV}(\sqrt{\Lambda_0}/10^{13}~{\rm GeV})^{1/2}$.
 Finally, we comment on how the inflationary paradigm,
BEH mechanism and phase transitions in the early Universe
get modified in the light of our findings. 
\end{abstract}


\maketitle

\section{Introduction}

 The cosmological constant problem can be stated 
follows~\cite{Weinberg:1988cp}. Vacuum fluctuations 
of any particle species contribute to the energy density of the vacuum as, 
\begin{equation}
 \rho_0 = \pm \int \frac{d^3p}{(2\pi)^3}\frac{1}{2}({p^2 + m^2})^{1/2}
\,,
\end{equation}
where $p$ is the momentum of a particle, $m$ its mass, 
and the positive (negative) sign stands for bosons (fermions). 
(In this article we use the natural units in which
the speed of light and the reduced Planck constant are set to unity,
$c=1$ and $\hbar = 1$.)
$\rho_0$ is a badly divergent quantity. 
It is often argued that a natural ultraviolet cutoff
is the Planck scale, $m_P = G_N^{-1/2} \simeq 1.22\times 10^{19}~{\rm GeV}$,
such that, 
\begin{equation}
 \rho_0 = \pm \frac{m_{\rm P}^4}{16\pi^2}
           \simeq \pm (3.4\times 10^{18}~{\rm GeV})^4
\,,
\label{rho0:matter}
\end{equation}
of equivalently, 
\begin{equation}
 \Lambda_0 = \pm \frac{\rho_0}{M_{\rm P}^2} 
           \simeq \pm 2.4\times 10^{37}~{\rm GeV}^2
\,,
\label{Lambda:natural}
\end{equation}
where $M_{\rm P} = (8\pi G_N)^{-1/2}\simeq 2.4\times 10^{18}~{\rm GeV}$
is the reduced Planck mass. Above the Planck scale,
an as of yet unknown quantum gravity theory is believed to
lead to finite dynamics. The observed value~\cite{Riess:1998,Perlmutter:1998}
is at least 120 orders of magnitude smaller, 
\begin{equation}
 \Lambda_{\rm obs} \leq 3H_0^2 \Omega_{\rm DE}
                   \simeq 5\times 10^{-84}~{\rm GeV}^2
\,,
\label{Lambda:observed}
\end{equation}
where we used the current value for the Hubble parameter,
$H_0\simeq 1.5\times 10^{-42}~{\rm GeV}$, and 
$\Omega_{\rm DE} \simeq  0.74$ is the dark energy contribution.

 The discrepancy between~(\ref{Lambda:observed}) and~(\ref{Lambda:natural})
is known as the cosmological constant problem. 
 There is a milder formulation of the problem, according to which 
supersymmetry takes care of the balance between the positive and 
negative contributions in~(\ref{Lambda:natural})
above the scale at which supersymmetry is restored.
In this case, $\rho_0\sim \pm m_{\rm susy}^4$, where 
$m_{\rm susy}$ is the energy scale above which supersymmetry is restored.
This works only for global supersymmetry however, since in local supersymmetry
(which also includes gravity) this balance does not necessarily hold. 

 Zeldovich has proposed~\cite{Zeldovich:1967gd} that the discrepancy
between the observed~(\ref{Lambda:observed}) and the theoretically
more natural value~(\ref{Lambda:natural}) could be explained
by a mechanism, where quantum matter fluctuations~(\ref{rho0:matter})
would cancel an initial geometric cosmological constant 
of Einstein's theory. Yet neither he, nor anyone else,
has been able to provide a satisfactory compensatory mechanism. 
In this article we present such a dynamical mechanism, 
which explains how an initially large and positive $\Lambda_0$ 
can be driven to zero by the gravitational backreaction induced by  
fermionic quantum fluctuations, such that today 
the effective cosmological term has a small, but nonvanishing, value.

 The cosmological constant has a long 
history~\cite{Weinberg:1988cp,RughZinkernagel:2000}
since it was in 1917 introduced by Einstein to his
general relativistic theory of gravitation.
Around the same time Nernst~\cite{Nernst:1916} observed that
quantum matter (light) fluctuations can contribute to the vacuum energy.
In 1968 the vacuum energy was experimentally observed~\cite{Boyer:1968uf}
based on the Casimir effect~\cite{Casimir:1948dh}.
Here we just briefly mention several attempts to solve
the cosmological constant problem. 
It is well known that (tree level) scalar potentials 
({\it e.g.} quintessence models~\cite{PeeblesRatra:2002}) 
cannot do the job because of the Weinberg
theorem~\cite{Weinberg:1988cp}. This theorem states
that one can always add a constant potential energy without dynamical
consequences to any scalar field potential which na\"ively drives
the field towards the value, where its potential energy vanishes.
Another possibility is ``shadow'' matter, which contributes 
with an opposite sign to the energy density as ordinary matter,
and couples only gravitationally. This requires introduction
of a lot of new and yet unobserved particle species,
and moreover has problems when perturbative gravity corrections
are taken into account~\cite{KaplanSundrum:2005}. 
The general trend in literature 
is either to resort to the anthropic principle~\cite{Weinberg:1988cp}, 
or to postulate new symmetries, an example being the symmetry
which relates positive and (as of yet unobserved) negative energy
states~\cite{Linde:1984ir,KaplanSundrum:2005,'tHooftNobbenhuis:2006}.

\bigskip

 When written in the approximation of a homogeneous and isotropic
space-time with the metric tensor, 
\begin{equation}
 g_{\mu\nu} = {\rm diag}(-1,a^2,a^2,a^2)
\,,
\label{metric tensor}
\end{equation}
the semiclassical Einstein equations can be written 
in the following FLRW form, 
\begin{eqnarray}
 H^2 &\equiv& \big(\frac{\dot a}{a}\Big)^2
    = \frac{8\pi G_N}{3}\rho + \frac{\Lambda_0}{3} - \frac{k}{a^2}
\label{FLRW:A}
\\
 \frac{\ddot a}{a} 
    &=& -\frac{4\pi G_N}{3}(\rho +3{\cal P})
        + \frac{\Lambda_0}{3}
\,,
\label{FLRW:B}
\end{eqnarray}
where $a$ denotes the scale factor, $\dot a = da/dt$, $\ddot a = d^2a/dt^2$,
$H$ is the Hubble parameter, $G_N$ the Newton constant,
$\Lambda_0$ the cosmological term, $\rho$ the energy density, 
${\cal P}$ the pressure, and $k$ is the curvature of spatial sections
of space-time. Here we take for simplicity $k=0$,
which is consistent with current observations~\cite{Spergel:2006}.
In the derivation of Eqs.~(\ref{FLRW:A}--\ref{FLRW:B}) one assumes 
that the matter stress-energy tensor has an ideal fluid form, 
%
%
and Eqs.~(\ref{FLRW:A}--\ref{FLRW:B}) are written in the fluid rest frame.
 
 We shall now calculate the stress-energy contribution
to the Einstein equations~(\ref{FLRW:A}--\ref{FLRW:B}) in the Yukawa theory. 
In doing so we shall resort to certain approximations, which are needed
for analytical calculations. 

The dynamics of the matter fields are described by the following 
model Yukawa theory,
\begin{eqnarray}
 S_{\rm Yu} = \int d^4 x \sqrt{-g} {\cal L}_{\rm Yu}
\,,
\label{Yukawa action}
\end{eqnarray}
with the Lagrangian,
\begin{eqnarray}
\sqrt{-g} {\cal L}_{\rm Yu} = 
 \sqrt{-g}
   \Big(\bar\psi i \nabla\!\!\!\!\slash \psi 
   -  m_0 \bar\psi\psi  
   - \frac12 g^{\mu\nu}(\partial_\mu\varphi)(\partial_\nu\varphi)
   - V_0(\varphi) 
   - \frac12 \xi_0 {\cal R}\varphi^2
   - f_0 \varphi \bar\psi\psi  
 \Big)
\,,
\label{Yukawa lagrangian}
\end{eqnarray}
where $\nabla\!\!\!\!\slash = e^\mu_b\gamma^b (\partial_\mu - \Gamma_\mu)$ 
denotes the covariant derivative acting on spinors, 
$e^\mu_b$ is the tetrad field, $\Gamma_\mu$ is 
the spin(or) connection, $m_0$ is the bare fermion mass,
$\psi$ and $\varphi$ denote the fermionic and scalar fields, respectively, 
$V_0 = (1/2)\mu_0^2 \varphi^2 + \lambda_0 \varphi^4/4!$
is the (tree level) scalar potential,
$\mu_0$ and $\lambda_0$ are the bare scalar mass and quartic self-coupling,
respectively,
${\cal R}$ is the Ricci curvature scalar, $g={\rm det}(g_{\mu\nu})$,
and $f_0$ is the Yukawa coupling. 

 Note that fermions appear quadratically in the action, so in principle 
they can be integrated out. In practice though,
this can be done only in a handful of gravitational backgrounds.
Since we are interested in solving 
the problem in a background given by a (positive or negative) cosmological
term plus matter contribution, we make the first of our crude 
approximations and assume that the gravitational
background is well approximated by (anti-)de Sitter space-time.
Below we comment on how good this approximation is. 
The problem of integrating fermions in de Sitter background
has recently been solved by Miao and Woodard~\cite{MiaoWoodard:2006},
whereby they neglected scalar field fluctuations.
This approximation is accurate to leading log,
where the log refers to $\ln(a)$, and $a$ is the scale factor of 
the Universe (see Eq.~(\ref{dS:scale factor}) below). The leading log
approximation is accurate in de Sitter background~\cite{Woodard:2005cw}.
Certain aspects of the work in Ref.~\cite{MiaoWoodard:2006} 
are inspired by an earlier work of 
Candelas and Raine~\cite{CandelasRaine:1975}.
Various quantum aspects of the Yukawa theory in de Sitter background
have formerly been studied in 
Refs.~\cite{GarbrechtProkopec:2006,MiaoWoodard:2005,ProkopecWoodard:2003},
and a new fermion mass generation mechanism has been
found~\cite{GarbrechtProkopec:2006}.

\section{de Sitter and anti-de Sitter spaces}
\label{de Sitter and anti-de Sitter spaces}

de  Sitter space is perhaps best viewed as a 4-dimensional
hyperboloid embedded into the five-dimensional Minkowski
space-time with the line element,
\begin{equation}
 ds_5^5 =  - dX_0 ^2 + dX_1^2 + dX_2^2 + dX_3^2 + dX_4^2
\label{dS:5 dim}
\,.
\end{equation}
The embedded hyperboloid of de Sitter space is shown in figure~\ref{figure 1},
and it is determined by 
\begin{equation}  
\label{dS:hyperboloid}
 - X_0^2 + X_1^2 + X_2^2 + X_3^2 + X_4^2  = \frac{1}{H^2}
\,,
\end{equation}
where $H$ denotes the Hubble parameter. 
The symmetry of de Sitter space,
$SO(1,4)$, is manifest in this embedding. One can define de Sitter
invariant distance functions, 
\begin{equation}
 Z(X;X^\prime) = H^2 \sum_{A,B=0}^4\eta_{AB}X_A X^\prime_B
               = 1 - \frac{1}{2}Y(X;X^\prime)
\,,\qquad
\eta_{AB} = {\rm diag}(-1,1,1,1,1)
\,.
\label{invariant distance}
\end{equation}
We shall use the following flat four-dimensional coordinates
(which cover 1/2 of de Sitter manifold),
\begin{eqnarray}
  X_0 &=& \frac{1}{H}\sinh(Ht)+\frac{H}{2}{\rm e}^{Ht}\|\vec x\|^2
  \,,\qquad 
            (-\infty < t < \infty)
\nonumber\\
 X_i &=& {\rm e}^{Ht}x_i
  \,\qquad 
            (-\infty < x_i < \infty)
 \,,\qquad
 (i=1,2,3)
\nonumber\\
 X_4 &=& \frac{1}{H}\cosh(Ht)-\frac{H}{2}{\rm e}^{Ht}\|\vec x\|^2
\,,
\label{dS:flat coordinates}
\end{eqnarray}
in which the metric tensor reduces to the form~(\ref{metric tensor}),
with the scale factor, $a={\rm e}^{Ht}$. 
When written in terms of conformal time $\eta$,
defined as $ad\eta = dt$, the metric tensor acquires the conformal form,
\begin{equation}
 g_{\mu\nu} = a^2 \eta_{\mu\nu}
\,,\qquad
 a = -\frac{1}{H\eta} \quad (\eta<0)
\,,\qquad
\eta_{\mu\nu} = {\rm diag}(-1,1,1,1)
\,.
\label{dS:scale factor}
\end{equation}
\begin{figure}[tbp]
\vskip -0.1in
\begin{center}
\epsfig{file=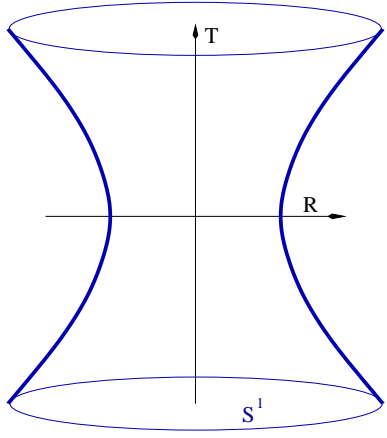,width=2.in}
\end{center}
\lbfig{figure 1}
\vskip -0.3in
\caption[fig0]{%
\small
The embedding of de Sitter space into a five dimensional flat space-time.
The vertical line corresponds to the time coordinate, $X_0=T$,
and the radial coordinate, $R=({X_1^2+X_2^2+X_3^2+X_4^2})^{1/2} = 1/H$.
At each point $(T,R)$ there is a unit 3-sphere, $S^3$
(which has in this figure collapsed to $S^1$).
}
\end{figure}
The invariant distance functions, $Z(X;X^\prime) \equiv z(x;x^\prime)$ 
and $Y(X;X^\prime) \equiv y(x;x^\prime)$, reduce in these coordinates to
the simple form, 
\begin{equation}
 z(x;x^\prime) = 1 - \frac 12 y(x;x^\prime)
\,,\qquad
 y(x;x^\prime) = aa^\prime H^2 \Delta x^2
\label{invariant distance:2}
\end{equation}
with $a=a(\eta) = -1/(H\eta)$,  
$a^\prime = a(\eta^\prime) = -1/(H\eta^\prime)$, 
and 
\begin{equation}
\Delta x^2(x;x^\prime) = -(|\eta-\eta^\prime|-i\epsilon)^2
                + \|\vec x-\vec x^\prime\|^2
\,, 
\label{Delta x}
\end{equation}
where (for later use)  we introduced an infinitesimal $\epsilon>0$. 
In these coordinates the curvature of spatial sections vanishes,
and thus they are also known as flat (Euclidean) coordinates, in which  
de Sitter space appears as uniformly expanding. 

For the integration of fermions we also need the fermion propagator,
\begin{equation}
 iS(x;x^\prime) = \langle x|
                  \frac{i}{\sqrt{-g}(i\nabla\!\!\!\!\slash - m)}
                  |x^\prime\rangle
\,,\qquad (m\equiv m_0+f_0\varphi)
\,.
\label{de Sitter propagator}
\end{equation}
This propagator can be solved in de Sitter background and in the 
approximation of a nearly constant $\varphi$ 
($\partial_\mu \varphi$ and higher order derivatives are neglected), 
and the solution can be written in the 
form~\cite{CandelasRaine:1975} \cite{MiaoWoodard:2006}, 
\begin{eqnarray}
  iS(x;x^\prime) = a\Big(i\nabla\!\!\!\!\slash + m\Big)
                    (aa^\prime)^{-\frac12}
                  \sum_{\pm} iS_{\pm}(y)\frac{1\pm\gamma^0}{2}  
\label{dS:fermion propagator}
\end{eqnarray}
where $i\nabla\!\!\!\!\slash 
  = a^{-\frac{D+1}{2}}i\gamma^\mu\partial_\mu a^{\frac{D-1}{2}}$,
and $iS_{\pm}=iS_{\pm}(y)$ are the following
de Sitter invariant scalar functions
(which contain no spinor structure), 
\begin{equation}
 iS_{\pm} = \frac{H^{D-2}}{(4\pi)^\frac{D}{2}}
            \frac{\Gamma\Big(\frac{D}{2}-1\mp i\frac{m}{H}\Big)
                    \Gamma\Big(\frac{D}{2}\pm i\frac{m}{H}\Big)}
                 {\Gamma\Big(\frac{D}{2}\Big)}
    \; {_2F_1}\Big(\frac{D}{2}-1\mp i\frac{m}{H},
                   \frac{D}{2}\pm i\frac{m}{H};
                   \frac{D}{2};
                   1 - \frac{y}{4}
              \Big)
\,,
\label{dS:fermion propagator:2}
\end{equation}
where $m=m(\varphi)$ is defined in Eq.~(\ref{de Sitter propagator}), 
$D$ denotes the number of space-time dimensions, ${_2F_1}$ 
is the hypergeometric function, and $\gamma^\mu$ are the Dirac
$\gamma$-matrices. With the $\epsilon$ pole prescription as 
in~(\ref{invariant distance:2}--\ref{Delta x}), one can show that the 
propagator~(\ref{dS:fermion propagator}--\ref{dS:fermion propagator:2}) 
indeed solves the Dirac equation in de Sitter space, 
\begin{equation}
\sqrt{-g}(i\nabla\!\!\!\!\slash - m)iS(x;x^\prime) 
          = i \delta^D(x-x^\prime)
\,,
\label{Dirac propagator equation}
\end{equation}
where $\delta^D(x-x^\prime)$
denotes the $D$-dimensional Dirac $\delta$-distribution
and $\varphi =$ const.

 Similarly to de Sitter space, anti-de Sitter space can be thought of
as the hyperboloid,
\begin{equation}  
 - X_0^2 - X_1^2 + X_2^2 + X_3^2 + X_4^2  = - R^2 \equiv - \frac{1}{H^2}
\label{AdS:hyperboloid}
\,,
\end{equation}
embedded into the five-dimensional flat space-time with the line element, 
\begin{equation}
 ds_5^5 =  - dX_0 ^2 - dX_1^2 + dX_2^2 + dX_3^2 + dX_4^2
\label{AdS:5 dim}
\,,
\end{equation}
which possesses an SO(2,3) symmetry. The graphical representation of 
anti-de Sitter space is the same as in figure~\ref{figure 1}, 
except that each point on the vertical axis
corresponds to a circle, and the points of constant radii,
$R=(X_2^2+X_3^2+X_4^2)^{1/2}$, correspond to two-dimensional 
spheres $S^2$. The definition of anti-de Sitter invariant 
distance functions is as in de Sitter space~(\ref{invariant distance}),
with the five dimensional metric given by,
 $\eta_{AB} = {\rm diag}(-1,-1,1,1,1)$.

The following coordinates, which cover a 1/2 of anti-de Sitter space,
are analogous to the flat de Sitter coordinates~(\ref{dS:flat coordinates}),
\begin{eqnarray}
  X_0 &=& \frac{1}{2u} + \frac{u}{2}\Big(R^2 + x_2^2 + x_3^2 - \eta^2\Big)
  \,\qquad 
            (-\infty < \eta < \infty, u>0)
\nonumber\\
 X_1 &=& Ru\eta
\nonumber\\
 X_i &=& Rux_i \qquad (i=2,3)
  \,\qquad 
            (-\infty < x_i < \infty)
\nonumber\\
 X_4 &=&  \frac{1}{2u} - \frac{u}{2}\Big(R^2 - x_2^2 - x_3^2 + \eta^2\Big)
\,,
\label{AdS:flat coordinates}
\end{eqnarray}
With the additional replacement, $u = 1/x_1$, 
these coordinates yield the conformal metric tensor,
\begin{equation}
 g_{\mu\nu} = a^2 \eta_{\mu\nu}
\,,\qquad 
   a = \frac{1}{Hx_1} \quad (x_1>0)
\,,\qquad
   H = \frac{1}{R}
\,.
\label{AdS:metric}
\end{equation}
One should not confuse $H$ in anti-de Sitter space
(where it denotes the inverse radius of curvature of the space)
with the Hubble parameter $H$ in de Sitter space. 
Note that unlike de Sitter space, which represents an expanding space time
in conformal flat coordinates, the conformal flat coordinates of 
anti-de Sitter space~(\ref{AdS:metric})
correspond to a static slicing with the space-time 
curved in one spatial direction, which we choose to be 
$x_1$. One can easily calculate the anti-de Sitter invariant distance function,
$\bar Z = -H^2\sum_{A,B}\eta_{AB}X_A X_B^\prime = 1+\bar Y/2$. 
Upon defining, $y(x;x^\prime) \equiv \bar Y(X;X^\prime)$
and $z(x;x^\prime) \equiv \bar Z(X;X^\prime)$, we find in these coordinates,
\begin{equation}
 z(x;x^\prime) = 1 + \frac 12  y(x;x^\prime)
\,,\qquad 
 y = aa^\prime H^2 \Delta x^2
\,,
\end{equation}
where here $a=a(x_1) = 1/(Hx_1)$,
 $a^\prime = a({x_1}^\prime) = 1/(H{x_1}^\prime)$ $(x_1,{x_1}^\prime >0)$
and $\Delta x^2$ is given in Eq.~(\ref{Delta x}).

 One can show that the anti-de Sitter fermion propagator
in the presence of an approximately constant scalar field 
reduces to the form~\cite{ProkopecRigopoulos:2006},
\begin{eqnarray}
  i \bar S(x;x^\prime) = a\Big(i\nabla\!\!\!\!\slash + m\Big)
                    (aa^\prime)^{-\frac12}
                  \sum_{\pm} i\bar S_{\pm}(y)\frac{1\pm i\gamma^1}{2}  
\,,
\label{AdS:fermion propagator}
\end{eqnarray}
where $i\bar S_{\pm}=i\bar S_{\pm}(y)$ 
are the anti-de Sitter invariant scalar functions,
%
\begin{equation}
 i\bar S_{\pm} = {\rm e}^{i(D-2)\pi/2}\frac{H^{D-2}}{(4\pi)^\frac{D}{2}}
            \frac{\Gamma\Big(\frac{D}{2}-1\mp \frac{m}{H}\Big)
                    \Gamma\Big(\frac{D}{2}\pm \frac{m}{H}\Big)}
                 {\Gamma\Big(\frac{D}{2}\Big)}
    \; {_2F_1}\Big(\frac{D}{2}-1\mp \frac{m}{H},
                   \frac{D}{2}\pm \frac{m}{H};
                   \frac{D}{2};
                   1 + \frac{y}{4}
              \Big)
\,,
\label{AdS:fermion propagator:2}
\end{equation}
and $m\equiv m_0 + f_0\varphi$.
Note that the two propagators 
(\ref{dS:fermion propagator}--\ref{dS:fermion propagator:2}) 
and~(\ref{AdS:fermion propagator}--\ref{AdS:fermion propagator:2}) 
are related by the analytic continuation,
\begin{equation}
 H\rightarrow iH
\,.
\label{analytic continuation}
\end{equation}
The extra phase in~(\ref{AdS:fermion propagator:2}) comes from 
the term in the propagator proportional to $y^{1-D/2}$.
This term gives rise to the $\delta$-function in the 
propagator equation~(\ref{Dirac propagator equation}), 
which induces an extra phase due to,  
$y = aa^\prime H^2\Delta x^2 \rightarrow -y$.
 The spinor structure of the 
two propagators is related such that the positive and negative energy
projectors are replaced as, $(1\pm \gamma^0)/2\rightarrow (1\pm i\gamma^1)/2$. 
The extra imaginary $i$ in the projectors, $(1\pm i\gamma^1)/2$,
is important, since it assures that   
the spinor structure drops out from the differential operator
acting on the propagator.

\section{Effective potentials}
\label{Effective potentials}

 When one uses the anti-de Sitter 
propagator~(\ref{AdS:fermion propagator}--\ref{AdS:fermion propagator:2})
to calculate the effective action by an analogous procedure 
as done in Refs.~\cite{CandelasRaine:1975} and~\cite{MiaoWoodard:2006},
one obtains the following (renormalised) effective potential, 
\begin{eqnarray}
 \bar V(\varphi) = V_{\rm tree}(\varphi)
                 +    \frac{H^4}{8\pi^2} 
              \bigg[
                   2\gamma_E\Big(\frac{m}{H}\Big)^2
                 + (\zeta(3)\!-\!\gamma_E)\Big(\frac{m}{H}\Big)^4
          \!+\! 2\int_{x_0}^{m/H}\!\!\! dx(x-x^3)\Big(\psi(1+x)+\psi(1-x)\Big)
              \bigg]
\,,\quad
\label{AdS:effective potential}
\end{eqnarray}
where,  
\begin{equation}
V_{\rm tree}(\varphi) = \frac{1}{2}\mu^2\varphi^2 
                 + \frac{\lambda}{4!}\varphi^4
                 - \frac{1}{2}{\cal R}\xi \varphi^2
\,,
\label{V tree}
\end{equation}
$\psi(z) = (d/dz)\ln(\Gamma(z))$ is the digamma function,  
and $\lambda$,  $\mu$ and $\xi$ are the renormalised parameters 
(in Ref.~\cite{MiaoWoodard:2006} the renormalisation 
has been chosen such that these parameters vanish), 
$\zeta(3) \simeq 1.202$ is the Riemann zeta function, and 
$\gamma_E \simeq 0.577$ is the Euler constant. 
 This effective potential is induced by fermionic fluctuations 
in the presence of a nearly constant scalar field in anti-de Sitter space. 
Since the integrand has simple poles at $x=2,3,4,..$, the lower limit of 
integration can be taken as $x_0=0$ for $m/H<2$, and $x_0=[m/H]+c_{AdS}$
when $m/H>2$, where $0<c_{AdS}<1$ is a constant, and 
$[\cdot]$ denotes the integer part. 
For $m/H>2$ the potential~(\ref{AdS:effective potential}) 
contains an unspecified constant, which has no dynamical relevance
(the dynamical relevance is associated with $-dV/d\varphi$). 

 Just like the propagators, the effective potentials are 
related by the analytic continuation~(\ref{analytic continuation}),
such that the de Sitter 
effective potential reads~\cite{MiaoWoodard:2006},
\begin{eqnarray}
  V(\varphi) = V_{\rm tree}(\varphi) 
             + \frac{H^4}{8\pi^2} 
               \bigg[
               \!-\!  2\gamma_E\Big(\frac{m}{H}\Big)^2
            \!+\! (\zeta(3)\!-\!\gamma_E)\Big(\frac{m}{H}\Big)^4
     \!-\!2\int_0^{m/H}\!\!\! dx(x+x^3)\Big(\psi(1\!+\!ix)+\psi(1\!-\!ix)\Big)
              \bigg]
\,.\quad
\label{dS:effective potential}
\end{eqnarray}
The large field limit of this effective potential is,
\begin{equation}
 V \simeq  -\frac{m^4}{8\pi^2}\ln\Big(\frac{m}{H}\Big)
\qquad (m\gg H)
\,.
\label{V:large m/H}
\end{equation}
 The potentials~(\ref{AdS:effective potential}--\ref{dS:effective potential})
determine the dynamics of scalar fields in de Sitter and anti-de Sitter
backgrounds. The dynamics in de Sitter space 
is given either by the corresponding Euler-Lagrange equations, 
\begin{equation}
 \partial_t^2 \varphi - \frac{1}{a^2}{\vec\partial}^{\,2} \varphi 
     + 3H \partial_t \varphi = -\frac{dV}{d\varphi} 
\,,
\label{dS:euler-lagrange}
\end{equation}
or by the Starobinsky stochastic 
theory~\cite{StarobinskyYokoyama:1994} \cite{Woodard:2005cw},
according to which
\begin{equation}
  \partial_t\varphi   = -\frac{1}{3H}\frac{dV}{d\varphi} + \xi\\
\,,
\label{dS:stochastic equation}
\end{equation}
where $\xi$ is the white noise
\begin{equation}
 \langle \xi(x)\xi(x^\prime) \rangle = \frac{H^3}{4\pi^2} \delta(t-t^\prime)
                                     \frac{\sin(Ha\|\vec x-\vec x^\prime\|)}
                                          {Ha\|\vec x-\vec x^\prime\|}
\,,
\label{white noise}
\end{equation}
generated in accelerating space-times as vacuum fluctuations 
get stretched beyond the Hubble radius, $1/H$. 
When the effects of quantum tunneling and stochastic fluctuations are
neglected, the field generally evolves in the direction of
decreasing potential, $dV/d\varphi<0$.

In anti-de Sitter space the dynamics is determined by the 
Euler-Lagrange equation, 
\begin{equation}
 \frac{1}{a^2}\partial_\eta^2 \varphi
         - \frac{1}{a^2}{\vec\partial}^{\,2} \varphi 
         + 2H \frac{1}{a}\frac{\partial}{\partial x_1} \varphi
                   = -\frac{d\bar V}{d\varphi} 
\label{AdS:euler-lagrange}
\end{equation}
where $\bar V$ is given in~(\ref{AdS:effective potential}).
Note that the ``damping'' (third) term in anti-de Sitter space
generates inhomogeneities and thus has a different meaning from
its de Sitter counterpart in Eq.~(\ref{dS:euler-lagrange}). 
Just like in de Sitter space, the Euler-Lagrange
equation~(\ref{AdS:euler-lagrange}) tells as that the field $\varphi$ is 
driven in the direction of decreasing potential (along which 
$-d\bar V/d\varphi$ is positive). 

 In the large field limit ($m=m_0+f_0\varphi\gg H$),
the effective potential~(\ref{dS:effective potential}) reduces to the
form that deceptively looks like the Coleman-Weinberg effective 
potential~\cite{ColemanWeinberg:1973,CandelasRaine:1975,MiaoWoodard:2006},
which exhibits the well known instability, according to which
fermion fluctuations drive the scalar field to infinity,
$\varphi\rightarrow \infty$, whereby $V\rightarrow -\infty$.
 This instability has been declared a problem.
As we shall argue, when interpreted correctly,
 this runaway feature is the cure.

\section{Stress-energy tensor}
\label{Stress-energy tensor}

\begin{figure}[htbp]
\begin{center}
\epsfig{file=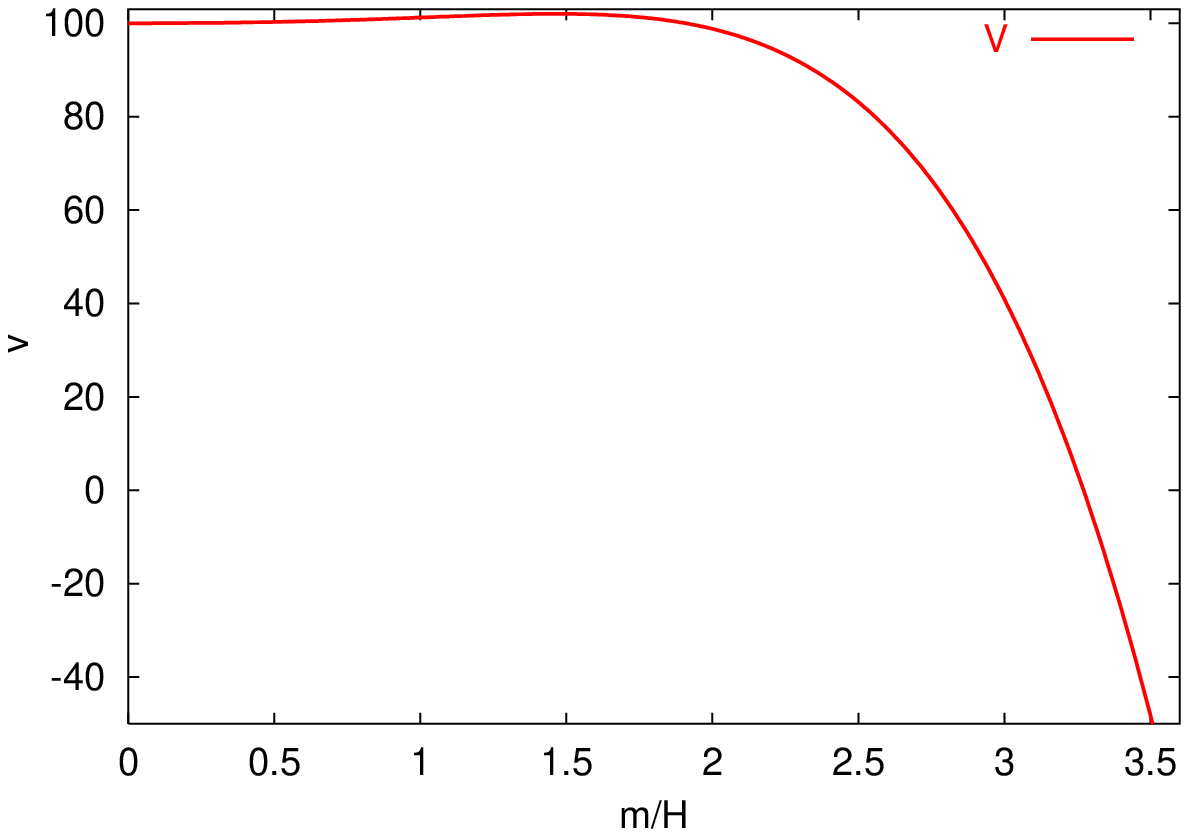, width=3.5in
       }
\epsfig{file=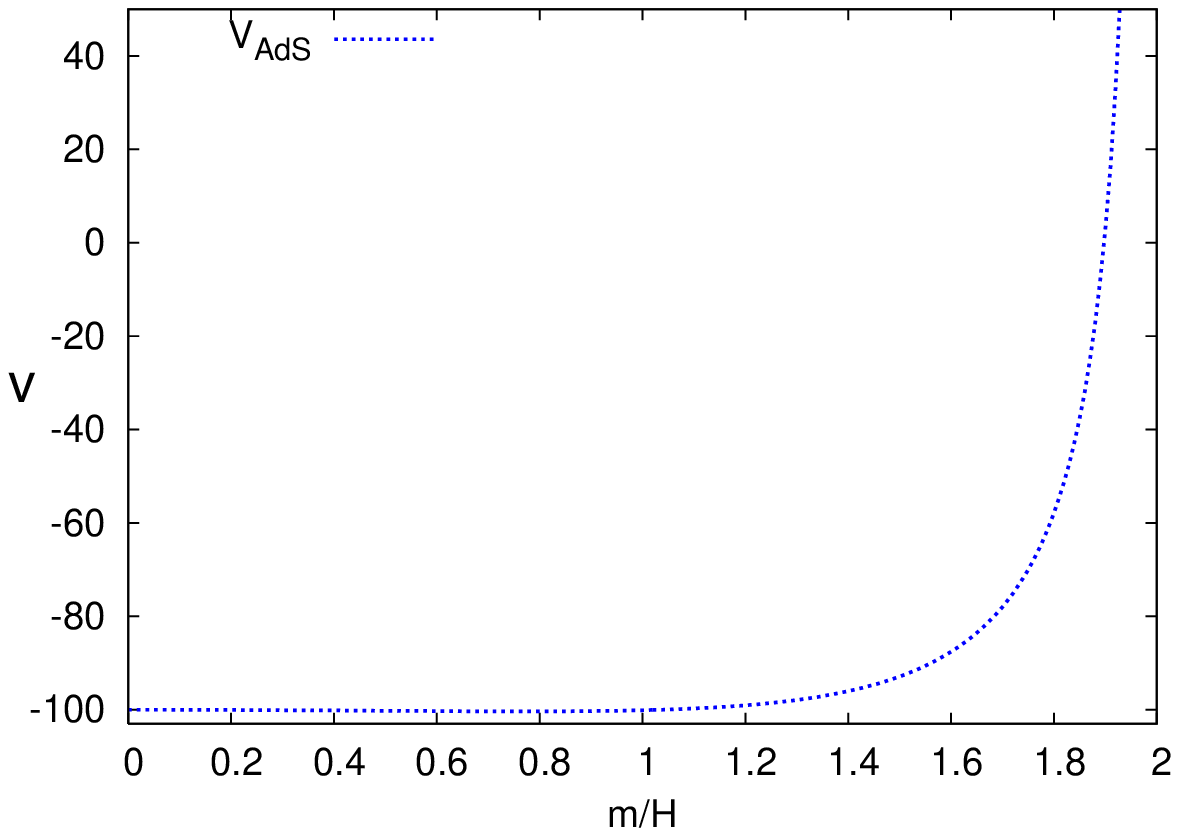, width=3.5in
       }
\end{center}
\vskip -0.3in
\lbfig{figure 2}
\caption{%
\small
{\it Na\"ive} potential energies, $v = v_0 + (16\pi^2/H^4)V_s$,
as a function of $m/H=(m_0+f_0\varphi)/H$ for de Sitter (left plot) 
and anti-de Sitter background space (right plot).
Note that, as $\varphi$ increases, the potential energy decreases (increases) 
in de Sitter (anti-de Sitter) space.
For definiteness, we chose $\xi =0 $,  $\mu = 0$, $\lambda =0 $
and $v(0) = 100\, (-100)$ in de Sitter (anti-de Sitter) space.
}
\vskip -0.in
\end{figure}
 We have so far said nothing about the back-reaction of
matter fields onto gravity, as indicated in
Eqs.~(\ref{FLRW:A}--\ref{FLRW:B}). To address this question,
we need to calculate the stress-energy tensor of the 
Yukawa theory~(\ref{Yukawa lagrangian}), which is obtained
by the variation of the action~(\ref{Yukawa action}) with respect to 
the metric tensor (or tetrad). The relevant part of the stress-energy 
tensor is (here we suppress terms, which involve space-time 
derivatives acting on scalar fields, and whose contribution
tend to be suppressed in accelerating space-times), 
\begin{eqnarray}
T_{\mu\nu} &=& 
    \frac{a}{2}{\rm lim}_{x\rightarrow x^\prime}
     \Big(
          i\gamma_{(\mu}\nabla_{\nu)} iS(x;x^\prime)
      -   iS(x;x^\prime) \overleftarrow{\nabla}^\prime_{(\mu}i\gamma_{\nu)}
     \Big)
 - \xi_0  G_{\mu\nu}\varphi^2
   + \frac{1}{2}\mu_0^2\phi^2 + \frac{\lambda_0}{4!}\phi^4
   + {\cal O}\Big((\partial\varphi)^2\Big)
\,,\qquad
\label{Yukawa stress energy}
\end{eqnarray}
where $\gamma_{(\mu}\nabla_{\nu)}$ denotes symmetrisation
with respect to $\mu$ and $\nu$, and 
$G_{\mu\nu} = {\cal R}_{\mu\nu} - \frac12 g_{\mu\nu}{\cal R}$
denotes the Einstein curvature tensor. 
Upon evaluating $T_{\mu\nu}$ in anti-de Sitter background and performing 
dimensional regularisation and renormalisation, 
one obtains~\cite{ProkopecRigopoulos:2006} ($m=m_0+f_0\varphi$), 
\begin{eqnarray}
\bar T_{\mu\nu} &=& - \bar V_s(\varphi) g_{\mu\nu}
   + {\cal O}\Big((\partial\varphi)^2\Big)
\nonumber\\
 \bar V_s(\varphi) 
 &=& V_{\rm tree}
\label{AdS:potential-rho}
\\
  &&\hskip -1.5cm
   +  \frac{H^4}{16\pi^2}
        \Bigg[
             (2\gamma_E\!-\!1)\Big(\frac{m}{H}\Big)^2
 \!+\!  \Big(\frac12 \!+\! 2\zeta(3)\!-\!2\gamma_E\Big)\Big(\frac{m}{H}\Big)^4
          + \Big(\frac{m}{H}\Big)^2\bigg(1\!-\!\Big(\frac{m}{H}\Big)^2\bigg)
           \bigg(
            \psi\Big(1\!+\!\frac{m}{H}\Big) + \psi\Big(1\!-\!\frac{m}{H}\Big)
           \bigg)
        \Bigg]
\,.
\nonumber
\end{eqnarray}
The de Sitter potential energy is, as expected, obtained by 
the analytic continuation~(\ref{analytic continuation}) 
of Eq.~(\ref{AdS:potential-rho}),
\begin{eqnarray}
  V_s(\varphi) 
 &=& V_{\rm tree}
\label{dS:potential-rho}
\\
  &&\hskip -1.5cm
  -  \frac{H^4}{16\pi^2}
        \Bigg[
             (2\gamma_E\!-\!1)\Big(\frac{m}{H}\Big)^2
 \!-\!  \Big(\frac12 \!+\! 2\zeta(3)\!-\!2\gamma_E\Big)\Big(\frac{m}{H}\Big)^4
          + \Big(\frac{m}{H}\Big)^2\bigg(1\!+\!\Big(\frac{m}{H}\Big)^2\bigg)
           \bigg(
            \psi\Big(1\!+\!i\frac{m}{H}\Big) + \psi\Big(1\!-\!i\frac{m}{H}\Big)
           \bigg)
        \Bigg]
\,,
\nonumber
\end{eqnarray}
which agrees with the result obtained in~\cite{MiaoWoodard:2006}.
We plot these two potential energies in figure~\ref{figure 2}. 
Note that the de Sitter potential energy shows instability and 
decays for large values of the argument, 
while the anti-de Sitter potential is stable. 
As the argument approaches, $m/H\rightarrow 2$,  anti-de Sitter potential
grows and diverges as, $\bar V_s\rightarrow+\infty$, 
such that the two curves cross.
For each point on the de Sitter curve, there is an anti-de Sitter curve
which crosses it. If they were both correct, the field dynamics would 
drive the field both towards larger values (along the de Sitter curve) 
as well as towards smaller values (along the anti-de Sitter curve).
Obviously, both cannot simultaneously correspond to the correct dynamics.

\begin{figure}[htbp]
\begin{center}
\epsfig{file=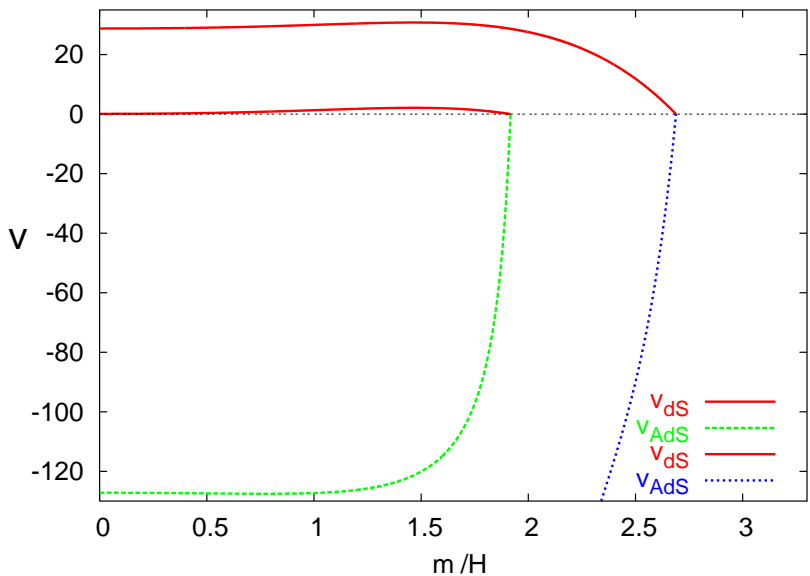, width=4.5in
       }
\end{center}
\vskip -0.3in
\lbfig{figure 3}
\caption{%
\small
Improved, but still na\"ive, potential energies,
$v_s = v_0 + (16\pi^2/H^4)V_s$,
as a function of $m/H=(m_0+f_0\varphi)/H$ 
for de Sitter and anti-de Sitter space-times.
Two (anti-)de Sitter curves are shown. The initial potential energy
for the de Sitter and anti-de Sitter case is chosen such that they match at
$V_{\rm eff} = 0$. 
As in
figure~\ref{figure 2}, we chose, $\xi =0 $,  $\mu = 0$, $\lambda =0$.
}
\vskip -0.in
\end{figure}
 This inconsistency is resolved by noting
that de Sitter potentials are valid for all $\rho_{\rm eff}\geq 0$,
while anti-de Sitter potentials are valid for $\rho_{\rm eff}\leq 0$, 
$\rho_{\rm eff} = 0 $ ($H=0$) being the limiting case, where 
\begin{equation}
 \rho_{\rm eff} \approx  V_{\rm eff} \equiv 
  \cases{V_s + M_{\rm P}^2\Lambda_0\,, &    
                for $V_{\rm eff} \geq 0$ and $\Lambda_0\geq 0$ \cr
         \bar V_s + M_{\rm P}^2\Lambda_0\,, & 
                    for $V_{\rm eff} \leq 0$ and $\Lambda_0\leq 0$\cr}
\,.
\label{Veff}
\end{equation}
Here $M_{\rm P}^2 = (8\pi G_N)^{-1} \simeq 2.4\times 10^{18}~{\rm GeV}$
denotes the reduced Planck mass. This leads to an improved
dynamics, two examples of which are shown in figure~\ref{figure 3}.
Note that the anti-de Sitter potential energy has simple poles 
at $m = nH$, where $n=2,3,4,..$ are the integers greater than or equal to 2,
such that, for example, the second anti-de Sitter curve in 
figure~\ref{figure 3} runs off to minus infinity
at $m = 2H$. The first reaction of a learned reader may be: 
if this picture were true, that would be disastrous, 
since that would imply that Minkowski 
space with $\Lambda_{\rm eff} = 0$ would be unstable, and decay into 
anti-de Sitter space, from where it would run either to some large 
and negative value of $\Lambda_{\rm eff}$ (when $m/H<2$ at $V_{\rm eff}=0$)
or even worse to $\Lambda_{\rm eff}\rightarrow -\infty$ 
(when $m/H>2$ at $V_{\rm eff} = 0$). 
This results in a very uncomfortable observation that 
an unfortunate field fluctuation or a tunneling event 
might take us to anti-de Sitter space, resulting in dire consequences.  
 There is one way we can be saved: if the field evolution slows down
dramatically as one approaches $\Lambda_{\rm eff} =0$ from above,
we would never reach anti-de Sitter.
As we shall argue, this is indeed what happens. 
Even better, we shall see that the na\"ive interpretation in 
figure~\ref{figure 3} is incorrect. In fact, the de Sitter and
anti-de Sitter curves do {\it not} continuously connect at
$\Lambda_{\rm eff}=0$, as suggested by figure~\ref{figure 3},
and the curves with $m>2H$ are in fact unphysical.
Moreover, $\Lambda_{\rm eff} = 0$ is the stable end-point of the evolution
for all $\Lambda_0\geq 0$. 
We shall also see that the probability to tunnel to anti-de Sitter
space is vanishingly small.

\section{Dynamical relaxation of the cosmological term}
\label{Dynamical relaxation of the cosmological term}

The true dynamics in de Sitter space are obtained by taking account 
of the matter back-reaction {\it via} the Einstein equation~(\ref{FLRW:A}).
When kinetic energies are neglected
with respect to the potential energy $V_{\rm eff}$ in~(\ref{Veff}),
which is justified in de Sitter and quasi-de
Sitter space-times (slow-roll regime of primordial inflation), 
the Friedmann equation~(\ref{FLRW:A}) reduces to,
\begin{equation}
 H^2 \simeq \frac{V_{\rm eff}}{3M_{\rm P}^2} 
         \equiv \frac 13\Big(\Lambda_0 +  \frac{V_s}{M_{\rm P}^2}\Big)
          \equiv \frac{\Lambda_{\rm eff}}3
\,,
\label{FLRW:A2}
\end{equation}
with the scalar field dynamics given by Eq.~(\ref{dS:euler-lagrange}) 
or by Eqs.~(\ref{dS:stochastic equation}--\ref{white noise}). 

 Taking account of the matter backreaction means simply that  
Eq.~(\ref{FLRW:A2}) should be understood as $H=H(\varphi)$. 
The true potential energy $V_{\rm eff}=V_{\rm eff}(\varphi)$
(or equivalently the effective cosmological term, 
$\Lambda_{\rm eff} = V_{\rm eff}/M_{\rm P}^2$)
is then given by the self-consistent solution of 
equations~(\ref{FLRW:A2}) and~(\ref{dS:potential-rho}).
 
 To get analytical insight into the field dynamics, 
observe that in the limit when $m/H = (m_0 + f_0\varphi)/H \gg 1$, 
the potential energy~(\ref{dS:potential-rho}) reduces to the form,
\begin{eqnarray}
  V_s(\varphi) 
 \simeq  V_{\rm tree} 
  -  \frac{H^4}{8\pi^2}
        \Bigg[\Big(\frac{m}{H}\Big)^4
                   \Big[\ln\Big(\frac{m}{H}\Big)-\frac14+\gamma_E-\zeta(3)\Big]
           + \Big(\frac{m}{H}\Big)^2
                   \Big[\ln\Big(\frac{m}{H}\Big)-\frac{5}{12}+\gamma_E\Big]
           + \frac{11}{120}
        \Bigg]
\,,
\label{dS:potential-rho:large}
\end{eqnarray}
where $ V_{\rm tree}$ is given in Eq.~(\ref{V tree}).
At a first sight the potential~(\ref{dS:potential-rho:large})
seems to be of the Coleman-Weinberg type~\cite{ColemanWeinberg:1973},
whereby $H$ is interpreted as a renormalisation scale.
Moreover, for a sufficiently large expectation value of the field,
the potential seems to be unstable and na\"ively drives the field
into anti-de Sitter space. Yet when Eq.~(\ref{FLRW:A2}) is 
taken account of, the picture completely changes.
To see this we insert~(\ref{dS:potential-rho:large})
into~(\ref{FLRW:A2}) to get, 
\begin{equation}
 H^2 = \frac{\Lambda_0}{3} 
         - \frac{m^4}{3\pi m_{\rm P}^2}\ln\Big(\frac{m}{H}\Big)
\,,
\qquad
\qquad (m\gg H)
\,,
\label{FLRW:A:approx}
\end{equation}
where $m_{\rm P} = G_N^{-1/2} \simeq 1.22\times 10^{19}~{\rm GeV}$
and we kept only the leading order term in~(\ref{dS:potential-rho:large}).
 If we want to solve this for $H$ or $\Lambda_{\rm eff}$ as a function 
of $m$, the following form of~(\ref{FLRW:A:approx}) is 
suitable for iterative procedure,
\begin{equation}
 H = m\exp\Big(-\frac{\pi m_{\rm P}^2\Lambda_0}{m^4} 
               + \frac{3\pi m_{\rm P}^2}{m^2}\frac{H^2}{m^2}
          \Big)
\,.
\label{H:iterative}
\end{equation}
When $H^2 \ll m^4/(3\pi m_{\rm P}^2)\simeq (\Lambda_0/3)/\ln(m/H)$
(which is a stronger condition than $m\gg H$), 
already the leading order iteration becomes a good approximation, 
\begin{equation}
 H = m\exp\Big(-\frac{\pi m_{\rm P}^2\Lambda_0}{m^4} 
         \Big)
\qquad \Big(H^2 \ll \frac{m^4}{3\pi m_{\rm P}^2}\Big)
\,.
\label{H:iterative:0}
\end{equation}
A careful look at~(\ref{H:iterative:0}) reveals something quite bizarre.
The potential energy corresponding to~(\ref{H:iterative:0}) is simply, 
\begin{eqnarray}
 V_s &=&  - \Lambda_0M_{\rm P}^2 + V_{\rm res}
\nonumber\\
V_{\rm res} 
    &=& 3m^2M_{\rm P}^2\exp\Big(-\frac{2\pi m_{\rm P}^2\Lambda_0}{m^4} 
                          \Big) 
\,,
\label{new potential}
\end{eqnarray}
which is of a completely different form than the perturbative-looking 
potential~(\ref{dS:potential-rho:large}).
Moreover, the potential~(\ref{new potential}) is nonperturbative 
with respect to both gravitational and Yukawa coupling constants.
Furthermore, one can show that the full solution of Eq.~(\ref{FLRW:A:approx})
is also nonperturbative with respect to $\Lambda_0$. Recall that,
after integrating the fermions in a nontrivial
background (de Sitter) space-time, we 
obtained a seemingly perturbative effective potential. 
Yet, when the matter backreaction onto gravity is taken account of, 
the resulting effective potential becomes nonperturbative both in
the Yukawa coupling, as well as in the gravitational coupling constants
(the latter is quite common in classical gravity).

\begin{figure}[htbp]
\begin{center}
\epsfig{file=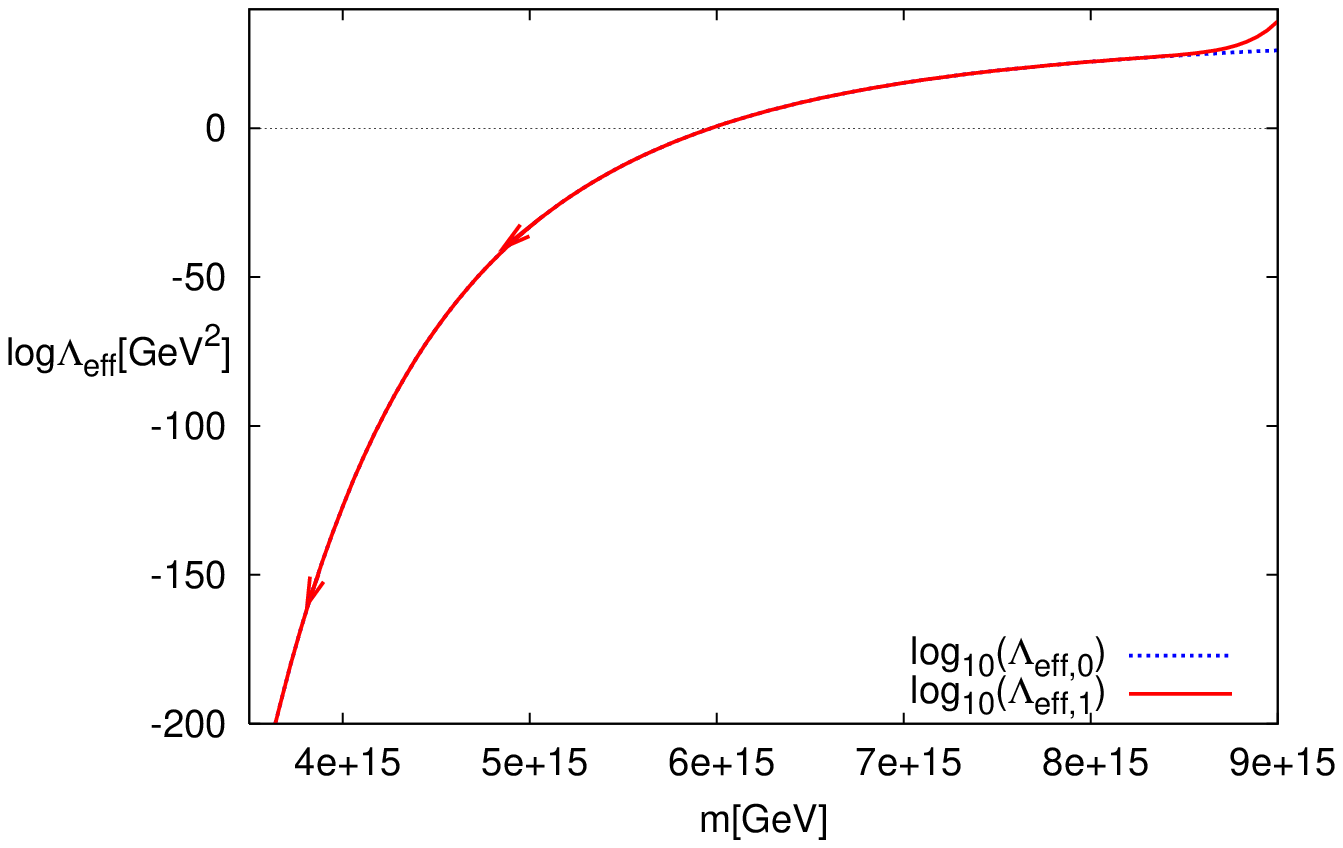, width=3.5in
       }
\epsfig{file=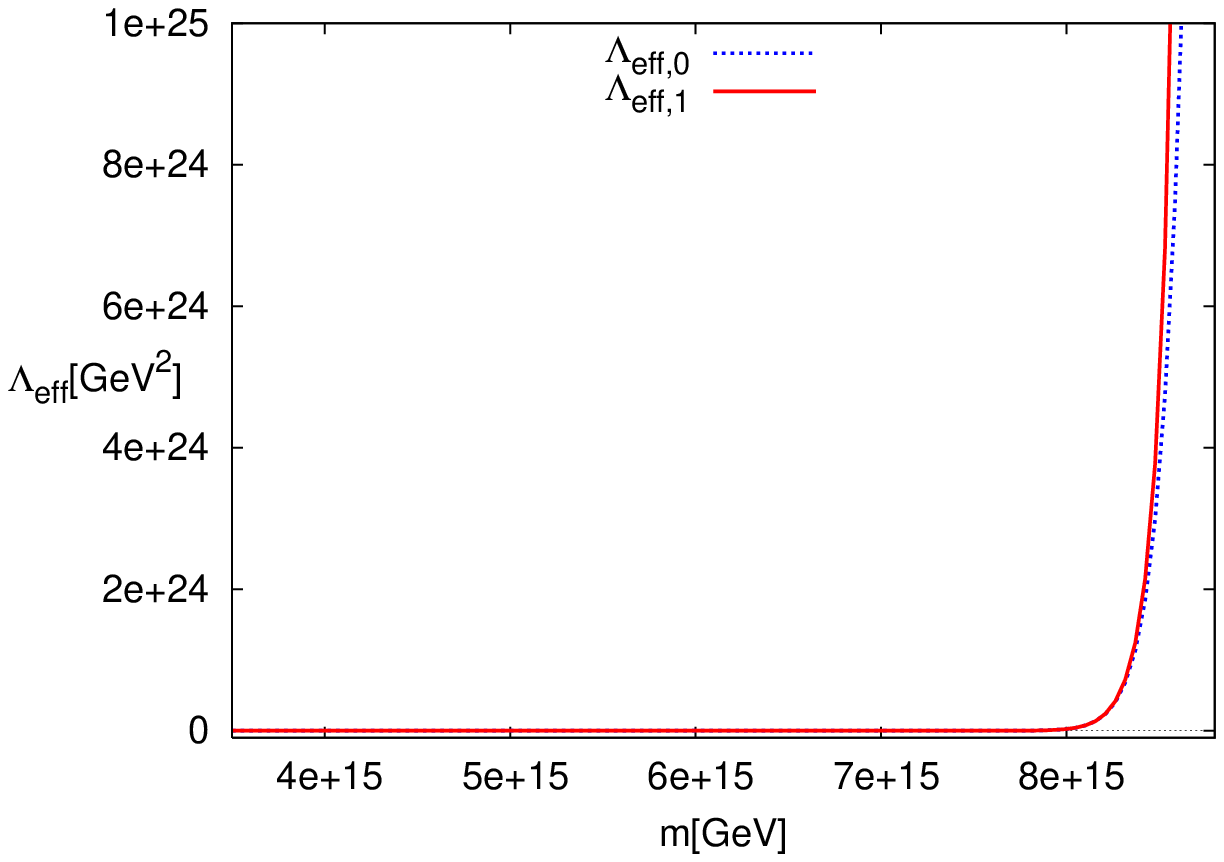, width=3.5in
       }
\end{center}
\vskip -0.3in
\lbfig{figure 4}
\caption{%
\small
The effective cosmological term $\Lambda_{\rm eff} = 3H^2$ 
as a function of $m = m_0 + f_0\varphi$. The left pannel shows
$\log_{10}(\Lambda_{\rm eff})$, the right pannel $\Lambda_{\rm eff}$.
As $m$ decreases, $\Lambda_{\rm eff}$ approaches {\it zero} faster than
exponentially. The {\it dashed (blue)} and {\it solid (red)} curves correspond
to the $0$th and $1$st order iterative solutions of Eq.~(\ref{H:iterative}),
respectively. From the left pannel we see that the effective cosmological term,
$\Lambda_{\rm eff} = 5\times 10^{-84}~{\rm GeV}^2$
in Eq.~(\ref{Lambda:observed}), 
corresponding to the dark energy density of the Universe today,
implies for the fermion mass,
$m=4.3\times 10^{15}~{\rm GeV}$ (where we chose 
$\sqrt{\Lambda_0}=10^{13}~{\rm GeV}$).
}
\vskip -0.in
\end{figure}
The effective cosmological term as inferred from Eq.~(\ref{H:iterative:0})
is,
\begin{equation}
 \Lambda_{\rm eff} =  3m^2\exp\Big(-\frac{2\pi m_{\rm P}^2\Lambda_0}{m^4} 
                              \Big)
\,,
\label{effective lambda}
\end{equation}
which approaches zero if $m$ approaches zero. Recall that $m=m_0+f_0\varphi$, 
such that this holds independently of the initial fermion mass $m_0$. 
What remains to be shown is that $m\rightarrow 0$ at late times. 
It suffices to show that the potential
energy~(\ref{new potential}) (at this level of 
approximation, $V=V_s$, {\it cf.} Eq.~(\ref{V:large m/H})) 
drives the system towards $m=0$ at late times,
where $\Lambda_{\rm eff} = 0$. Indeed, since within 
our approximations, 
$-V^\prime = -V_s^\prime \propto - m^\prime = -f_0 <0$
($V^\prime \equiv dV/d\varphi$), the force pulls 
the field towards the origin at $m = 0$, where both the potential
energy $V_s$ and $H^2 = \Lambda_{\rm eff}/3$ exhibit essential singularity. 
Visually this can be seen in figure~\ref{figure 4}, in which we plot
the effective cosmological term~(\ref{effective lambda}) as a function of
$m$. The arrows in figure~\ref{figure 4} indicate the direction along which 
the field evolves. To conclude, we found that the cosmological constant problem
is solved in the Yukawa theory plus Einstein's gravity in the following sense:
independently of the initial value of the cosmological term $\Lambda_0$, 
as long as it is positive (and smaller than the Planck scale), 
it is dynamically driven to zero,
\begin{equation}
 \Lambda_{\rm eff} \;\stackrel{m\rightarrow 0}{\longrightarrow}\; 0
\,.
\label{Lambda=0}
\end{equation}
Hence $\Lambda_{\rm eff} = 0$ is an {\it attractor} for 
{\it any positive} initial cosmological constant
$\Lambda_0>0$ in a Yukawa theory coupled to gravity. 
This is true for any Yukawa theory whose tree level couplings  
$\mu$, $\lambda$ and $\xi$ are such that the scalar field
does not get stuck forever in a false vacuum.

Next we address the question of tunneling from de Sitter 
to anti-de Sitter space.
The tunneling probability, $P_{\rm tunnel}$, can be estimated in terms of 
the Euclidean action $S_E$ as follows, 
\begin{eqnarray}
 P_{\rm tunnel} &\sim& {\rm e}^{-S_E}
\nonumber\\
 S_E &=& \int d^4 x_E \sqrt{g_E} M_{\rm P}^2\Lambda_{\rm eff}
\,.
\label{tunneling+euclidean action}
\end{eqnarray}
Working in closed coordinates (which cover whole de Sitter space), 
\begin{equation}
  ds_E^2 = \frac{1}{H^{2}\sinh^2(\eta_E)}\Big(d\eta_E^2 + d\chi^2 
          + \sin^2(\chi) \big(d\theta^2 + \sin^2(\theta)d\phi^2\big) \Big)
\,\quad(-\infty\leq \eta_E\leq\infty,\; 0\leq\chi\leq \pi)
\,,
\label{closed coordinates}
\end{equation}
we have $\sqrt{g_E} = H^{-4}\sin^2(\chi)\sin(\theta)/\sinh^4(\eta_E)$ 
such that the Euclidean action~(\ref{tunneling+euclidean action}) becomes,
\begin{eqnarray}
 S_E &=& \frac{2\pi m_{\rm P}^2}{H^2}
           \sinh^4\Big(\frac{Ht_E}{2}\Big)
                  \bigg(\sinh^2\Big(\frac{Ht_E}{2}\Big)+\frac32\bigg)
\,,
\label{Euclidean action}
\end{eqnarray}
where $t_E$ is the physical Euclidean time, which determines 
the instanton duration, and it is defined as,
${\rm e}^{2\eta_E}-1 = 1/\sinh^2(Ht_E/2)$. 
Note that the action~(\ref{Euclidean action})
 diverges as $t_E\rightarrow \infty$, implying 
that the tunneling is strictly speaking forbidden.
For a Hubble time instanton, $t_E\sim 1/H$,
Eq.~(\ref{Euclidean action}) yields,
\begin{eqnarray}
 S_E \sim \frac{m_{\rm P}^2}{H^2}
\,,
\label{Euclidean action:2}
\end{eqnarray}
such that the probability for Hubble time instantons is
suppressed as, $P_{\rm tunnel}\sim \exp(-m_{\rm P}^2/H^2)$, which is
so tiny that it is indistinguishable from zero today. 
A typical time interval that anti-de Sitter instantons can exist 
(for which $S_E\sim 1$) 
can be estimated from~(\ref{Euclidean action}) to be, 
$t_E \sim (m_{\rm P}H)^{-1/2} \simeq 2\times 10^{11}~{\rm GeV}^{-1}
        \sim 10^{-13}~{\rm sec}$. The question of the role of 
stochastic fluctuations is under 
investigation~\cite{ProkopecRigopoulos:2006}.

\bigskip

 Having established that de Sitter space is stable against rolling and
tunneling into anti-de Sitter space, we now comment on 
what happens if the Universe begins with a negative cosmological term
$\Lambda_0$.
The relevant potential energy curve is shown in figure~\ref{figure 3}
(the lower left ({\it dashed green}) curve). The field begins 
at some $\Lambda_0<0$, and since the potential is concave (curved upwards), 
the field fluctuates around $m = 0$, 
as given by Eq.~(\ref{AdS:euler-lagrange}). Just like in de Sitter 
space, the potential energy plotted in figure~\ref{figure 3} is na\"ive,
and the full account of the potential energy (when field fluctuations 
are suppressed) can be obtained by the self-consistent solution 
of Eq.~(\ref{AdS:potential-rho}) and the corresponding Friedmann equation
which can be obtained from Eq.~(\ref{FLRW:A}) by analytic continuation,
\begin{equation} 
 H^2 \equiv \Big(\frac{1}{a^2}\frac{da}{dx_1}\Big)^2
       = - \frac{\Lambda_0}{3}
         - \frac{\bar V_s(\varphi)}{3 M_{\rm P}^2} 
\,,
\end{equation}
where here $H=1/R$ is the inverse radius of curvature
of anti-de Sitter space, and 
we approximated the energy density by the corresponding 
anti-de Sitter potential energy~(\ref{AdS:potential-rho}), 
$\rho\simeq \bar V_s$. For a large curvature radius, 
$R=H^{-1}\gg \sqrt{3/|\Lambda_0|}$, 
the potential~(\ref{AdS:potential-rho}) can be expanded around the 
singular point $m = 2H$ (which is approached whenever
$|\Lambda_0|\ll m_{\rm P}^2$)
to obtain (we set $\mu,\lambda,\xi=0$), 
\begin{eqnarray}
 \bar V_s(\varphi) 
   =  \frac{3H^4}{4\pi^2}
        \Big(\frac{1}{2-m/H} + \frac 83\zeta(3) - \frac92\Big)
\qquad \big(H^2 \ll \Lambda_0/3\big)
\,.
\label{AdS:potential-rho:m=2H}
\end{eqnarray}
Just like in de Sitter space, the self-consistent solution 
in anti-de Sitter space exhibits two
branches. The effective potential at the upper
near-Minkowski branch (where $H\approx 0$) can be approximated by,
\begin{equation}
  \bar V_s \simeq -\Lambda_0 M_{\rm P}^2 - \frac 34 m^2 M_{\rm P}^2 
\,,
\label{AdS:upper branch}
\end{equation}
such that 
\begin{equation}
 \Lambda_{\rm eff} = - 3H^2 \simeq  - \frac{3}{4}m^2
\,,
\label{AdS:upper branch:2}
\end{equation}
which behaves like an inverted harmonic oscillator with 
a negative mass term~({\it cf.} Ref.~\cite{GuthPi:1982}),
\begin{equation}
 m_{AdS}^2 \simeq   - \frac 32 f_0^2 M_{\rm P}^2 
\,\qquad  (m_0=0,\;H^2 \ll -\Lambda_0/3)
\,.
\label{AdS:induced mass}
\end{equation}
In this analysis we have also assumed, $m\ll (m_{\rm P}^2|\Lambda_0|)^{1/4}$.
These results can be interpreted as follows. 
If the Universe begins with a negative initial $\Lambda_0$
and at the lower branch, then it is dynamically driven towards
the stable point at, $m=0, \Lambda_{\rm eff} = \Lambda_0$, which is
the true potential minimum. If, on the other hand, 
the Universe begins at the upper branch~(\ref{AdS:upper branch}), 
it will exhibit an inverted harmonic oscillator instability
with the mass~(\ref{AdS:induced mass}) and 
quickly descent to the lower branch and eventually end up again at
the stable point, $\Lambda_{\rm eff} = \Lambda_0$. 
In this sense there is no problem with stability of Minkowski space.

\section{Dark energy}
\label{Dark energy}

 Next we address the question whether the dark energy of 
the Universe~\cite{Riess:1998,Perlmutter:1998}, whose density is
\begin{equation}
 \Omega_{\rm DE} \equiv \frac{\rho_{\rm DE}}{\rho_{\rm cr}} = 0.74 \pm 0.04
\,,
\end{equation}
(the 3 year WMAP data~\cite{Spergel:2006} with a Hubble parameter prior
results in, $\Omega_{\rm DE} = 0.76{+0.04 \atop -0.06}$)
or equivalently 
\begin{equation}
  \Lambda_{\rm eff, DE}  \simeq 5\times 10^{-84}~{\rm GeV}^2 
\,,
\end{equation}
can be explained within our model, 
where,
$\rho_{\rm cr} = 3H_0^2 M_{\rm P}^2 \simeq 4\times 10^{-47}~{\rm (GeV)^4}$
is the critical energy density today.
To see that, we recast Eq.~(\ref{H:iterative:0})
in the following form suitable for iterations,
\begin{equation}
 m^4 = \frac{\pi m_{\rm P}^2\Lambda_0} {\ln\big(m/H\big)}
\,\qquad \Big(H^2\ll \frac{m^4}{3\pi m_{\rm P}^2}\Big)
\,,
\label{m-H}
\end{equation}
from where we obtain,
\begin{equation} 
  m = 4.3\times 10^{15}
          \Big(\frac{\sqrt{\Lambda_0}}{10^{13}~{\rm GeV}}\Big)^{1/2}~{\rm GeV}
\,.
\label{heaviest fermion=DE}
\end{equation}
The choice $\Lambda_0 = (10^{13}~{\rm GeV})^2$ is the natural scale
of primordial inflation~\cite{Starobinsky:1980te,Guth:1980zm}.
Note that $m\propto \Lambda_0 ^{1/4}\simeq H_I^{1/2}$ 
($H_I$ is the Hubble parameter during primordial inflation)
changes rather slowly, as the scale of primordial inflation changes.
Even though $m$ in Eq.~(\ref{heaviest fermion=DE}) is of the GUT scale,
the curvature of the effective potential~(\ref{new potential})
today is of the order the Hubble parameter today, $d^2 V/d\varphi^2\sim H_0^2$,
which is tiny. 
The mass~(\ref{heaviest fermion=DE}) is our prediction for the heaviest fermion
which acquires mass {\it via} the mechanism presented here. 
There may be heavier fermions in Nature, which couple to 
a (grand-unified) Higgs field~\cite{Higgs:1964}, 
{\it e.g.} to a field with a negative mass term and a positive quartic 
coupling, but they do not contribute to $\Omega_{\rm DE}$. Indeed,
the residual vacuum energy of all Higgs fields
and other scalar and condensate fields is already contained in $\Lambda_0$, 
such that the dynamical compensation mechanism presented here
works for the residual vacuum energies of any Brout-Englert-Higgs (BEH)
mechanism~\cite{EnglertBrout:1964}~\cite{Higgs:1964} and phase 
transition, which takes place in the early Universe. 
What is special about the scalar field responsible for our mechanism, 
is that it must not couple to gauge and other scalar fields 
which would result in an upward sloped effective potential.

What remains to be shown is that,
from the moment when the energy density becomes dominated by 
the residual fermionic energy density $V_{\rm res}$ in~(\ref{new potential}),
the Universe enters a slow roll regime. We thus need to estimate
the rate of change of the Hubble parameter. From Eq.~(\ref{H:iterative:0})
we infer,
\begin{eqnarray} 
 \frac{\dot H}{H}
   \simeq \frac{4\pi m_{\rm P}^2 \Lambda_0}{m^4} \frac{\dot m}{m}
\,.
\label{HdotH}
\end{eqnarray}
In the slow roll regime we have, $\dot\varphi \simeq -V^\prime/(3H)$.
This and Eq.~(\ref{new potential}) allow us to estimate $\dot m/m$,
such that~(\ref{HdotH}) gives, 
\begin{eqnarray} 
  \frac{\dot H}{H^2}
   \simeq - f_0^2 \frac{4\pi m_{\rm P}^6 \Lambda_0^2}{m^{10}}
\,.
\label{HdotH:2}
\end{eqnarray}
Slow roll regime requires, $|\dot H/H^2 |\ll 1$, which means that 
\begin{eqnarray} 
   f_0 \ll f_{\rm cr}=\frac{m^{5}}{\sqrt{4\pi} m_{\rm P}^3 \Lambda_0}
       \simeq 2.4\times 10^{-6}
            \Big(\frac{\sqrt{\Lambda_0}}{10^{13}~{\rm GeV}}\Big)^\frac12
\,.
\label{slow roll}
\end{eqnarray}
This is a rather small value, but still corresponds to a modest fine tunning.
The electron Yukawa is sufficiently small, $f_e\simeq 2\times 10^{-6}$, 
to marginally satisfy the bound~(\ref{slow roll}). 

 From the rate of change of dark energy, 
$\rho_{\rm DE} \simeq V_{\rm res}$, which is 
defined in Eq.~(\ref{new potential}), 
\begin{equation}
 \frac{\dot\rho_{\rm DE}}{\rho_{\rm DE}}
     = -3H(1+w_{\rm DE})
     = - 8\pi f_0^2 \frac{m_{\rm P}^6\Lambda_0^2H}{m^{10}}
\,,
\label{DE-dot}
\end{equation}
%
%
%
we immediately infer,
\begin{equation}
 \frac{\sqrt{\Lambda_0}}{f_0^2}
     = \frac{1}{12\pi^{3/2}}\frac{m_{\rm P}}{1+w_{\rm DE}}
        \bigg[\ln\Big(\frac{\pi m_{\rm P}^2\Lambda_0}{H^4}\Big)\bigg]^{5/2}
    \simeq \frac{1\times 10^{24}~{\rm GeV}}{1+w_{\rm DE}}
\,.
\label{DE-dot:2}
\end{equation}
When the slow roll condition~(\ref{slow roll}) is met, 
$w_{\rm DE}$ gets close to $-1$, as it should.
In evaluating~(\ref{DE-dot:2}) we took, $\Lambda_0^{1/2}=10^{13}~{\rm GeV}$. 
The {\it r.h.s.} changes by about $10\%$ for each order of magnitude change 
in the scale of primordial inflation, $\Lambda_0^{1/2}$. 
Since the current observational bounds on $w_{\rm DE}$,
$-1.2<(w_{\rm DE})_{\rm obs}\leq -0.8$~\cite{Seljak:2004xh} and 
$(w_{\rm DE})_{\rm obs} = -1.06{{+0.13}\atop{-0.08}}$~\cite{Spergel:2006},
do not resolve $w_{\rm DE}$ from $-1$, 
we cannot yet use~(\ref{DE-dot:2}) to constrain
the fundamental parameters, $f_0$ and $\Lambda_0$, of the theory.

Next we estimate the number of e-foldings and the time
before the Universe reaches the singular point
at $\Lambda_{\rm eff} = 0$ (flat Minkowski space). 
The number of e-foldings can be estimated 
as ($V^\prime = dV/d\varphi$), 
\begin{equation}
 N = \int H dt = -\frac{3}{f_0}\int \frac{H^2}{V^\prime}dm
   = \frac{m^6}{6f_0^2m_{\rm P}^4\Lambda_0}   
\,.
\end{equation}
Since we do not know the value of $f_0^2/\sqrt{\Lambda_0}$, we cannot yet 
calculate $N$. Based on the slow-roll bound~(\ref{slow roll}) we can
however estimate
the minimum number of e-foldings before the Universe hits the singularity 
at $m=0$, 
\begin{equation}
 N \geq N_{\rm cr} \equiv \frac{m^6}{6f_{\rm cr}^2m_{\rm P}^4\Lambda_0}   
     =   \frac 23 \ln\Big(\frac{m}{H}\Big)
     \simeq 88
\,.
\end{equation}
The amount of time left before the Universe reaches the singular point
can be estimated as, 
\begin{equation}
 t = \int dt = -\frac{3}{f_0}\int \frac{H}{V^\prime}dm
   = \frac{\pi^{5/4}}{4f_0^2}\frac{\Lambda_0^{1/4}}{m_{\rm P}^{3/2}}
      \int_{u_0}^\infty \frac{du}{u^{9/4}}{\rm e}^u
       \rightarrow \infty
\,,
\label{time left}
\end{equation}
with $ u_0 = (\pi m_{\rm p}^2\Lambda_0)/m^4 = \ln\big(m/H\big)  \simeq 132$. 
Hence, the singular point $\Lambda_{\rm eff}=0$ is reached
only at an infinite future, which protects us from the singularity.

 Flat Minkowski space is thus the singular point of Yukawa theory,
which is reached only at future infinity. If that point were ever reached,
strictly speaking one would loose all information about the initial state,
since all universes with an initial $\Lambda_0>0$ converge to that point,
representing a new type of information loss. Figure~\ref{figure 3} is 
thus incorrect in the following sense. If one ever reaches
the point $\Lambda_{\rm eff}=0$, $m=0$, any information about 
the initial $\Lambda_0$ would be lost, such that it makes no sense to
continuously connect de Sitter and anti-de Sitter curves, as 
indicated in figure~\ref{figure 3}. For a negative $\Lambda_0$,
the evolution in anti-de Sitter space is trivial,
in the sense that the effective cosmological term does not change,
$\Lambda_{\rm eff}=\Lambda_0$. If for some reason 
the Universe starts at the upper near-Minkowski
branch~(\ref{AdS:upper branch}),
$\Lambda_{\rm eff} \approx 0$, then it will rapidly descent
to $\Lambda_{\rm eff} = \Lambda_0<0$, as explained at the end of 
Section~\ref{Dynamical relaxation of the cosmological term}. 
This is really an instability associated with anti-de Sitter space.  
Minkowski space is the space with $\Lambda_0=0$, and it is thus fully stable.

When viewed as a function of the fermion mass $m$, as $m$ becomes smaller and
smaller,  the singular point $\Lambda_{\rm eff}=0$ is approached faster than
exponentially, implying that at late times trajectories 
with different initial $\Lambda_0$ rapidly bunch up, 
and as time goes on it becomes more difficult to resolve
the original $\Lambda_0$ of the Universe. 
Since our theory contains essentially 
three free parameters ($\Lambda_0, f_0$ and $m$), to resolve 
$\Lambda_0$ and $f_0$, in addition to $\Omega_{\rm DE}$ and $w_{\rm DE}$, 
one has to measure $dw_{\rm DE}/dt$, 
which is planned to be measured in the near future.

\section{Discussion}
\label{Discussion}

A dynamical mechanism for relaxation of the cosmological term is
presented in section~\ref{Dynamical relaxation of the cosmological term}
by making use of the Yukawa theory coupled to Einstein's gravity.
The mechanism is generic however and should apply to any theory, 
which in Minkowski space exhibits an effective scalar potential
with instability or runaway behavior.
When the matter backreaction to Einstein's gravity is taken into account, 
the scalar field is dynamically driven towards the attractor at
$\Lambda_{\rm eff} = 0$, for any initial positive cosmological term 
below the Planck scale, $\Lambda_0\ll m_{\rm P}^2$, and  
for any reasonable tree level coupling parameters. 
If $\Lambda_0<0$, the dynamics 
is trivial, and $\Lambda_{\rm eff} =\Lambda_0$. 

 Related work on Yukawa theory does not offer an explanation
for why $\Lambda_{\rm eff} \approx 0$ today. 
For example, in Ref.~\cite{CandelasRaine:1975}
we read, ``there is no particle creation in de Sitter space.'' 
In references which discuss the Coleman-Weinberg-type of 
effective theories, it is universally claimed that the effective
Yukawa theories pose a problem, because of the instability 
of the effective scalar potential induced by the radiative effects of fermions.
 
 An early attempt to attribute dark energy to the quantum fermionic
fluctuations in expanding space-times dates a couple of years 
back~\cite{GarbrechtProkopec:2004}. The attempt failed since at that time 
the authors were not aware of
Ref.~\cite{CandelasRaine:1975} and did not properly
calculate the contribution of fermionic fluctuations to the stress-energy 
tensor in de Sitter and other expanding backgrounds.

 It is worth noting that in the light of our dynamical relaxation 
mechanism for $\Lambda_{\rm eff}$,
the problem of constructing quasi-stable local minima with
a small but positive effective cosmological term,
an example being the recent construction within the context of string 
theory~\cite{KachruKalloshLindeTrivedi:2003}, loses its main motivation.

 The next question is whether it is reasonable 
to assume that the effective scalar potentials
in de Sitter and anti-de Sitter spaces are dominated
by the Yukawa coupling to fermions. 
It is well known that coupling to scalars contributes positively
to the scalar potential energy~\cite{ColemanWeinberg:1973,CandelasRaine:1975}
(in the latter reference, replace $m^2$ by $\lambda \varphi^2$).
It is less known that gauge fields contribute also positively
to the effective potential~\cite{ProkopecTsamisWoodard:2006}, such that 
for the dynamical mechanism to be operative as 
advocated, it must be that the coupling of scalars to fermions 
dominates over the coupling to scalars and gauge fields
at large scalar field expectation values at least for one scalar field.
If the heaviest fermion of that kind has a mass, 
$m\simeq 4.3\times 10^{15}~{\rm GeV}
 (\sqrt{\Lambda_0}/10^{13}~{\rm GeV})^{1/2}$,
and the Yukawa coupling not larger than the electron Yukawa, 
then our mechanism can explain the dark energy of 
the Universe, which makes up about $74\%$ of the energy density 
of the Universe, and which has a negative equation of state,
$w_{\rm DE} \in (-1.14,-0.93)$~\cite{Spergel:2006}.

The history of the Universe gets revised within the Yukawa
theory~(\ref{Yukawa action}--\ref{Yukawa lagrangian}) as follows. 
The Universe begins with an initial positive cosmological term,
$\Lambda_0>0$, and with a scalar field vacuum expectation value close to zero.
The latter can be achieved, for example, with a positive scalar mass term
or quartic self-coupling. The initial cosmological term 
is not just of geometric origin, but it also comprises 
contributions of vacuum fluctuations of all matter fields.
Even though the precise value of
$\Lambda_0$ is not important, if we want to realise
primordial inflation within our model, then
$\Lambda_0 \sim (10^{13}~{\rm GeV})^2$. 
If the initial $\Lambda_0$ differs significantly from
$(10^{13}~{\rm GeV})^2$, than one or 
a series of phase transitions in the pre-inflationary Universe
can eventually generate a value suitable for primordial inflation. 
What is important is that the last slow roll regime of the theory corresponds
to a scale, $\sqrt{\Lambda_0}\sim 10^{13}~{\rm GeV}$, since then the amplitude
of cosmological perturbations will correspond to the measured value.
Since vacuum fluctuations of fermions in de Sitter background generate 
the effective potential~(\ref{dS:effective potential}), 
the field rolls driven by, $-dV/d\varphi>0$.
If the potential happens to be trapped at a small value of the field by 
{\it e.g.} a positive scalar mass term, the field
eventually tunnels to a value above which $-dV/d\varphi>0$.
One gets a slow roll inflation and enough of e-foldings 
when the condition, $f_0\leq 10^{-3}$, is met
(note that this condition is milder than the slow roll condition
today~(\ref{slow roll})), such that 
the Universe's homogeneity, isotropy, flatness, age and size problems 
are solved in the usual way~\cite{Guth:1980zm}.
Our inflationary model neither suffers from the usual
runaway problem, presumed to be present in inflationary models
with the Coleman-Weinberg potentials~\cite{AlbrechtSteinhardt:1982},
nor it suffers from the naturalness problem related to the usual
fine tunning of the zero potential energy at the end of inflation. 

 After the end of inflation, the field decays,
the Universe reheats, and the potential energy is approximately given 
by~(\ref{dS:potential-rho:large}) and~(\ref{V:large m/H}), where 
$H$ is now dominated by the radiation contribution, and can be viewed as 
a thermal (renormalisation) scale. 
The nature of the potential changes again
in the matter era at a redshift, $z\simeq 0.3$, when dark energy starts 
dominating, and the Universe enters a new 
slow roll regime, driven by the effective potential~(\ref{new potential}). 
While our model does not really solve the coincidence problem
(why are the densities of dark energy and matter so similar today?),
we point out that the required fermion mass~(\ref{heaviest fermion=DE}),
whose fluctuations yield the correct amount of dark energy,
is of the GUT scale, which is well motivated by particle physics. 
If during some earlier epoch in radiation
or matter era the vacuum energy starts dominating
(this may happen for example after a Higgs-like phase transition, or 
at the time of chiral condensate formation),
then again~(\ref{new potential}) becomes the relevant potential,
driving dynamically $\Lambda_{\rm eff}$ toward smaller values.
This means that phase transitions mediated by scalar fields
or some other condensates, which may change the vacuum energy and thus 
the effective cosmological term, do not anymore pose a problem,
since any vacuum energy gets eventually compensated.
 Even though it is not necessary to change the standard electroweak BEH
mechanism~\cite{EnglertBrout:1964} \cite{Higgs:1964}, 
the Higgs potential may still happen to have a nonvanishing
negative mass-squared term and a positive quartic 
self-coupling. A positive quartic coupling is not any more required,
since the potential ultimately gets stabilised through the Yukawa
couplings to the Standard Model fermions by the mechanism presented here.
Since in this modified mass generation mechanism the Higgs mass 
would be of the order or smaller than the Hubble parameter today,
$m_H\sim H_0\sim 10^{-33}~{\rm eV}$, 
the Higgs particle would not be seen at the LHC or any future accelerator
experiments. If the quartic self-coupling of the Higgs field
is chosen to be strictly zero, 
then there is no hierarchy problem associated with large Higgs field 
radiative corrections (gauge fields radiative corrections still contribute).
Furthermore, the issue of the electroweak vacuum stability~\cite{Sher:1988},
and more generally the question whether this modified electroweak
mass generation mechanism is consistent with all accelerator experiments, 
requires further investigation.

Finally, there are many possible improvements to this work:
one can relax the assumption that the background 
space-time is (anti-)de Sitter, and work with quasi-(anti-)de Sitter 
space and more general expanding space-times;
one can study how gradient corrections affect the results, {\it etc.}
\vskip -2.6cm

 \section*{Acknowledgements}

 I wish to thank Richard Woodard for infinite patience when explaining to me
the how-to-do's of quantum field theories in curved space-times, and for 
a long term collaboration, which has been the main inspiration of this work. 
I would like to thank my advisor Robert Brandenberger for introducing me to
the cosmological constant problem. I thank to Gerasimos Rigopoulos for 
collaboration on the project, for carefully checking the manuscript, 
and for making important suggestions on how to improve it.
I thank Tomas Janssen for useful suggestions.


\end{document}